\documentclass[amsmath,amssymb,aps,prl,twocolumn,showpacs,superscriptaddress]{revtex4-1}
\usepackage{graphicx}
\usepackage{dcolumn}
\usepackage{bm}
\usepackage{xr}
\makeatletter
\usepackage{hyperref}
\usepackage{xcolor}
\usepackage{gensymb}
\hypersetup{
  colorlinks   = true,
  urlcolor     = black,
  linkcolor    = blue,
  citecolor   = blue
}

\begin{document}

\title{Collective stiffening of soft hair assemblies}

\author{Jean-Baptiste Thomazo}
\affiliation{Laboratoire Jean Perrin, Institut Biologie Paris-Seine, CNRS-UMR 8237, Sorbonne University, 4, place Jussieu, 75005 Paris, France}
\affiliation{Nestl\'{e} Dairy Center, Rue d'Orival, 14100 Lisieux, France}
\author{Eric Lauga}
\affiliation{Department of Applied Mathematics and Theoretical Physics, University of Cambridge, Cambridge CB3 0WA, United Kingdom}
\author{Benjamin Le R\'{e}v\'{e}rend}
\affiliation{Nestl\'{e} Research Center, Route du Jorat, 1000, Lausanne, Switzerland}
\author{E. Wandersman}
\email[]{elie.wandersman@sorbonne-universite.fr}
\affiliation{Laboratoire Jean Perrin, Institut Biologie Paris-Seine, CNRS-UMR 8237, Sorbonne University, 4, place Jussieu, 75005 Paris, France}
\author{A. M. Prevost}
\email[]{alexis.prevost@sorbonne-universite.fr}
\affiliation{Laboratoire Jean Perrin, Institut Biologie Paris-Seine, CNRS-UMR 8237, Sorbonne University, 4, place Jussieu, 75005 Paris, France}

\date{\today}

\begin{abstract}
Many living systems use assemblies of soft and slender structures whose deflections allow them to mechanically probe their immediate environment. In this work, we study the collective response of artificial soft hair assemblies to a shear flow by imaging their deflections. At all hair densities, the deflection is found to be proportional to the local shear stress with a proportionality factor that decreases with density. The measured collective stiffening of hairs is modeled both with a microscopic elastohydrodynamic model that takes into account long range hydrodynamic hair-hair interactions and a phenomenological model that treats the hair assemblies as an effective porous medium. While the microscopic model is in reasonable agreement with the experiments at low hair density, the phenomenological model is found to be predictive across the entire density range.
\end{abstract}

\maketitle

\begin{figure}[ht]
\centering
\includegraphics[width=0.48\textwidth]{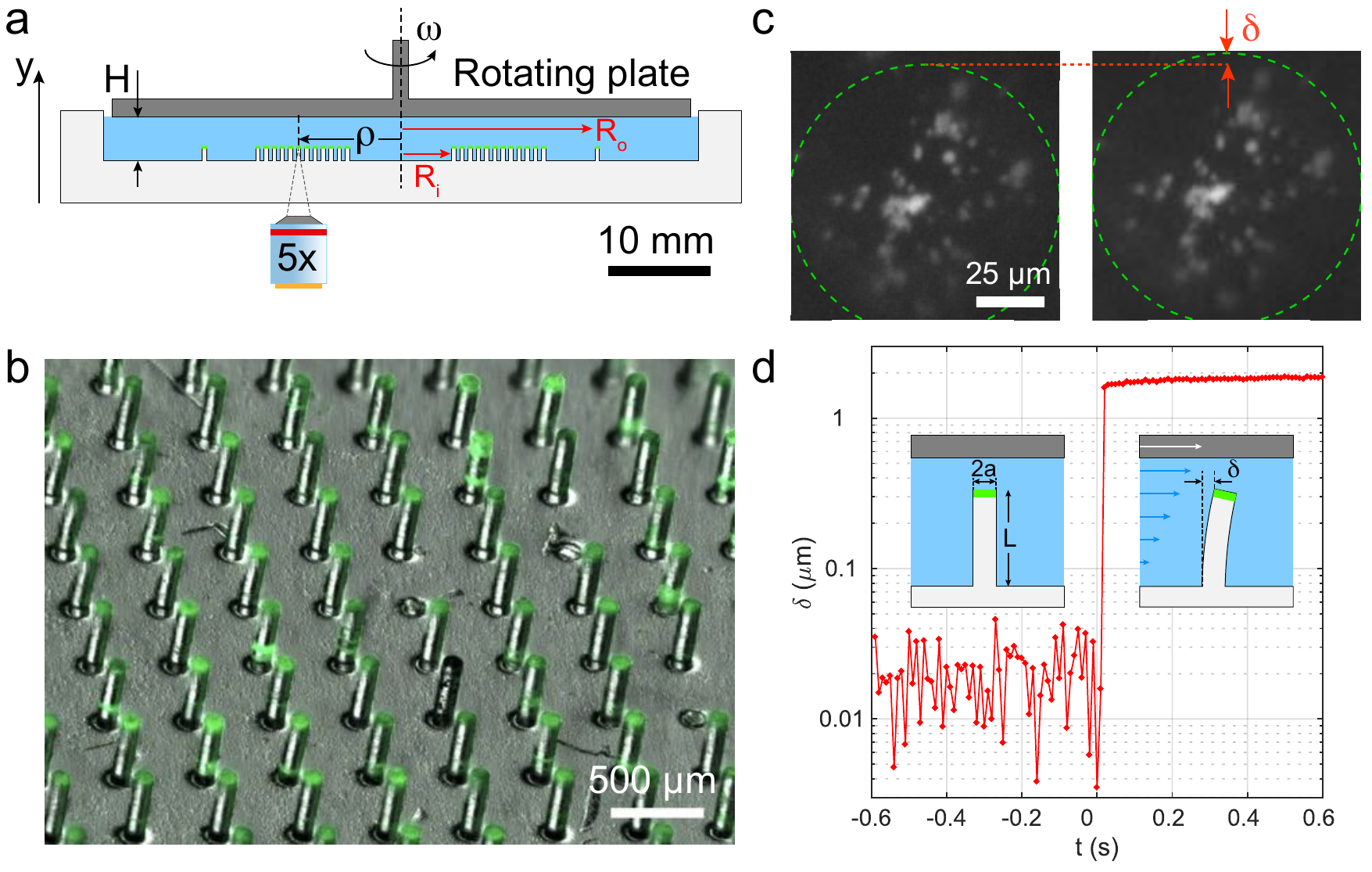}
\caption{(a) Sketch of the experimental setup (side view along a diameter). (b) Composite image of a PDMS substrate with a density of pillars $n \approx 4$~mm$^{-2}$ obtained with fluorescence imaging. The green color indicates the presence of green fluorescent particles on the pillar's tips. (c) Typical fluorescence snapshots of a pillar's summit at rest (left panel) and in steady flow (right panel). The green dashed line circles delimit the perimeter of the pillar. (d) Semi-log plot of the displacement $\delta$ of a single pillar versus time during a typical experiment after a sudden start of the rheometer's tool at $t = 0$~s. Inset: sketches of sectional views of a pillar at rest (left panel) and subject to a steady shear liquid flow (right panel).}
\label{fig:fig1}
\end{figure}

Many living systems use assemblies of soft and slender structures to mechanically probe their immediate environment. In animals, their deflection triggers the response of mechanosensitive nervous endings embedded at their base that convert mechanical stresses into a neural response~\citep{delmas2011molecular}. Rodents in particular use their facial whiskers to localize spatially objects and discriminate their texture by contact~\citep{arabzadeh2016vibrissal}. Spiders and crickets have legs covered with hairs that can detect minute changes in an air flow~\citep{bathellier2005viscosity,dangles2006ontogeny}. Fish have a lateral line that consist of an assembly of hair structures (neuromasts) allowing them to orient themselves in a flow and detect the presence of preys and predators in their vicinity~\citep{chagnaud2008measuring}. The human tongue itself is covered with filiform papillae that are deformed during food mastication and participate in in-mouth texture perception~\citep{ghom2008textbook,BookShimizu2012,doty2015handbook}. At smaller length scales, that of cells, slender structures (primary cilia) play an important role in mechanosensation processes~\citep{malicki2017cilium}. Given such ubiquity, numerous biomechanical models have been developed to predict the deformation of elongated structures when submitted to contact~\citep{quist2014modeling,boubenec2012whisker} and viscous stresses \citep{Venier1994,Lauga2016,du2019dynamics}. To test these models within simplified frameworks, artificial hairy systems have been used. These usually consist in slender pillars anchored to a substrate, whose deflections are monitored optically under controlled stresses~\citep{claverie2017whisker,axtmann2016investigation,wexler2013bending}.\\
\indent The deflection of an isolated pillar in a viscous flow has been successfully predicted using elastohydrodynamics~\citep{du2019dynamics,Lauga2016,ThomazoJRSI2019}. In biological systems however, hairs are usually densely packed and their mechanical behavior is likely subjected to hydrodynamic interactions. Indeed, the presence of a  pillar in a flow disturbs the velocity field around it and produces a long-range flow perturbation. When two pillars are sufficiently far apart, these perturbations decay sufficiently fast and hydrodynamic interactions are negligible. As pillars get closer, hydrodynamic interactions play an increasing role and modify the  deflection of pillars. Understanding mechanotransduction processes of hair assemblies thus requires to take into account the hydrodynamic couplings. Such interactions have been explored in the context of cell locomotion~\citep{lauga2009hydrodynamics}, fish schools~\cite{weihs1973hydromechanics} and bird flocks~\cite{higdon1978induced}, but are barely studied for anchored and passive fibers assemblies.\\ 
\indent In a previous work~\citep{ThomazoJRSI2019}, we have shown that deflections of an isolated elastomeric pillar submitted to a shear flow are proportional to the local shear stresses. In the context of in-mouth texture perception, we concluded that filiform papillae could act as sensitive stress sensors. In this Letter, we extend our biomimetic approach to assemblies of pillars at varying surface densities. We probe experimentally and theoretically how the density of pillars changes their collective deflections.\\ 
\indent We used a minimal biomimetic setup sketched in Fig.~\ref{fig:fig1}a (\textit{see}~\cite{ThomazoJRSI2019} and Supplemental Material for details). Briefly, it consists in mimicking soft hair assemblies with a pool made of an elastomer, whose bottom is decorated with cylindrical pillars of diameter $2a=100~\mu$m and length $L = 435\pm7~\mu$m. The pool is placed at the bottom of a rheometer (MCR~302, Anton Paar) whose PP40 planar rotating tool is used to impose the flow. This rheometer has built-in fluorescence microscopy capabilities allowing to image at 100~frames/s (fps) the pillars tips through the optically transparent pool. This is done thanks to fluorescent particles embedded at the tips. Soft hair assemblies were fabricated using micro-milling and elastomer molding techniques. Two types of patterns were drilled on the same mold, a first one consisting of isolated holes serving as references, and a second one comprising a square pattern of densely distributed holes of meshsize $d$. Hair assemblies were obtained by pouring in the molds a liquid PDMS (PolyDimethylSiloxane, Sylgard 184, Dow Corning, USA) -- crosslinker mixture, followed by a curing in an oven (12 hours, $T=65\degree$C) and unmolding. Figure~\ref{fig:fig1}b shows an example of a resulting substrate imaged using a macroscope with fluorescence imaging capabilities. The Young's modulus of the PDMS elastomer was measured to be $E=2.7\pm0.8$~MPa. Solutions of glycerol (Sigma-Aldrich) mixed in Millipore deionized water at different concentrations were used. Their dynamic viscosities $\eta$ were measured with the rheometer operating in a plate-plate geometry. Pillar tips deflections $\delta$ were measured~\cite{ThomazoJRSI2019} by correlating images of a pillar in its deformed state with a reference image where the pillar is at rest (Fig.~\ref{fig:fig1}c). From the 2D correlation function, we extract the maximum displacement $\delta$ with a spatial resolution of about 20~nm (Fig.~\ref{fig:fig1}d).\\
\indent Experiments were carried out as follows. First, a substrate with a given number density of pillars $n=1/d^2$ was positioned on the rheometer's base. Accurate determination of the gap $H$ (\textit{i.e.} distance from the base of the pool to the lower surface of the rotating plate) was determined. The pool was then filled with the liquid. The rotating plate was finally brought $550~\mu$m above the pillars summits (yielding $H=1$~mm) and set in motion at a constant angular velocity $\omega$, causing pillars deflection. Each experiment consisted of 11 successive 10~s long measurements. The first one was performed without any flow to provide an unperturbed reference state, while the 10 subsequent measurements were done with increasing $\omega$ distributed on a logarithmic scale. Their values were chosen so that $\delta$ ranges from 1 to 10~$\mu$m. The analysis of the displacements was performed in the steady state regime, yielding $\delta$ as a function of the shear stress $\sigma = \eta \dot{\gamma}$ with $\dot{\gamma} = \rho \omega/H$ the shear rate and $\rho$ the radial coordinate of the pillar (Fig.~\ref{fig:fig1}a). Given the error on the gap $H$ thickness (typically 20~$\mu$m, \cite{ThomazoJRSI2019}), the relative error on $\sigma$ is about 2\%. Over the whole range of shear rates $\dot{\gamma}$, the Reynolds number $Re = \rho L^2 \dot{\gamma} /\eta$ varies from $10^{-4}$ to $10^{-1}$, and thus the  flow remains laminar in all cases. For each density $n$, 6 independent experiments were carried out to measure the deflections of 3 different isolated pillars and 3 pillars in the dense region.     

\begin{figure}[h]
\centering
\includegraphics[width=0.45\textwidth]{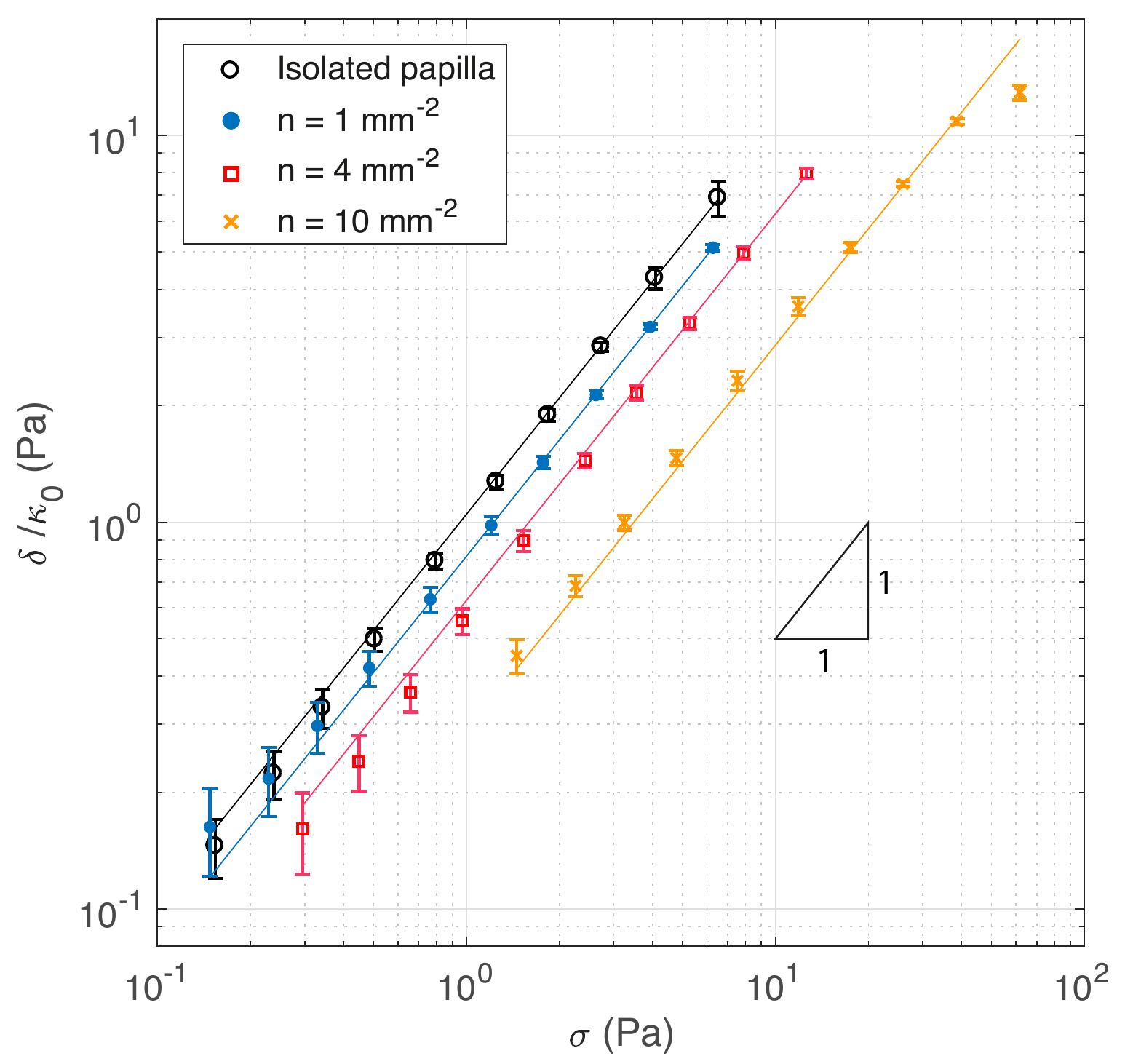}
\caption{Log-log plot of the normalized displacement, $\delta/\kappa_0$, of a single pillar's tip in steady state versus the local shear stress $\sigma$ for 4 different pillars number densities, $n=0.04$~mm$^{-2}$ (referred to as an isolated pillar, diamonds), $n=1$~mm$^{-2}$ (circles), $n=4$~mm$^{-2}$ (squares) and $n=10$~mm$^{-2}$ (stars). For these experiments, the angular velocity $\omega$ was varied and different pillar's radial positions $\rho$ and viscosities $\eta$ were used ($\rho=5.6, 7.0, 7.5$~mm and $\eta=0.458, 0.320, 0.468$~Pa.s for $n=1, 4, 10$~mm$^{-2}$ respectively). Solid lines are linear fits.}
\label{fig:fig2}
\end{figure}

As shown previously~\cite{ThomazoJRSI2019}, in steady state, the maximum displacement of an isolated pillar $\delta_0$ increases linearly with $\sigma$ (Fig.~\ref{fig:fig2}, diamonds), in agreement with the model of~\cite{Lauga2016} and writes
\begin{equation}
\delta_0 = K_0\frac{L^5}{a^4E}\frac{\eta \rho \omega}{H} = \kappa_0 \sigma
\label{eq:laugamodel1}
\end{equation}
\noindent with $K_0$ a numerical factor whose value is determined experimentally. Note that depending on the substrate, $\kappa_0$ can vary from 0.5 to 5 $\mu$m/Pa due to slight sample to sample variations of $L$ and $E$. This explains why we used composite pools consisting of two regions with isolated and dense pillars.

\begin{figure}[h]
\centering
\includegraphics[width=0.45\textwidth]{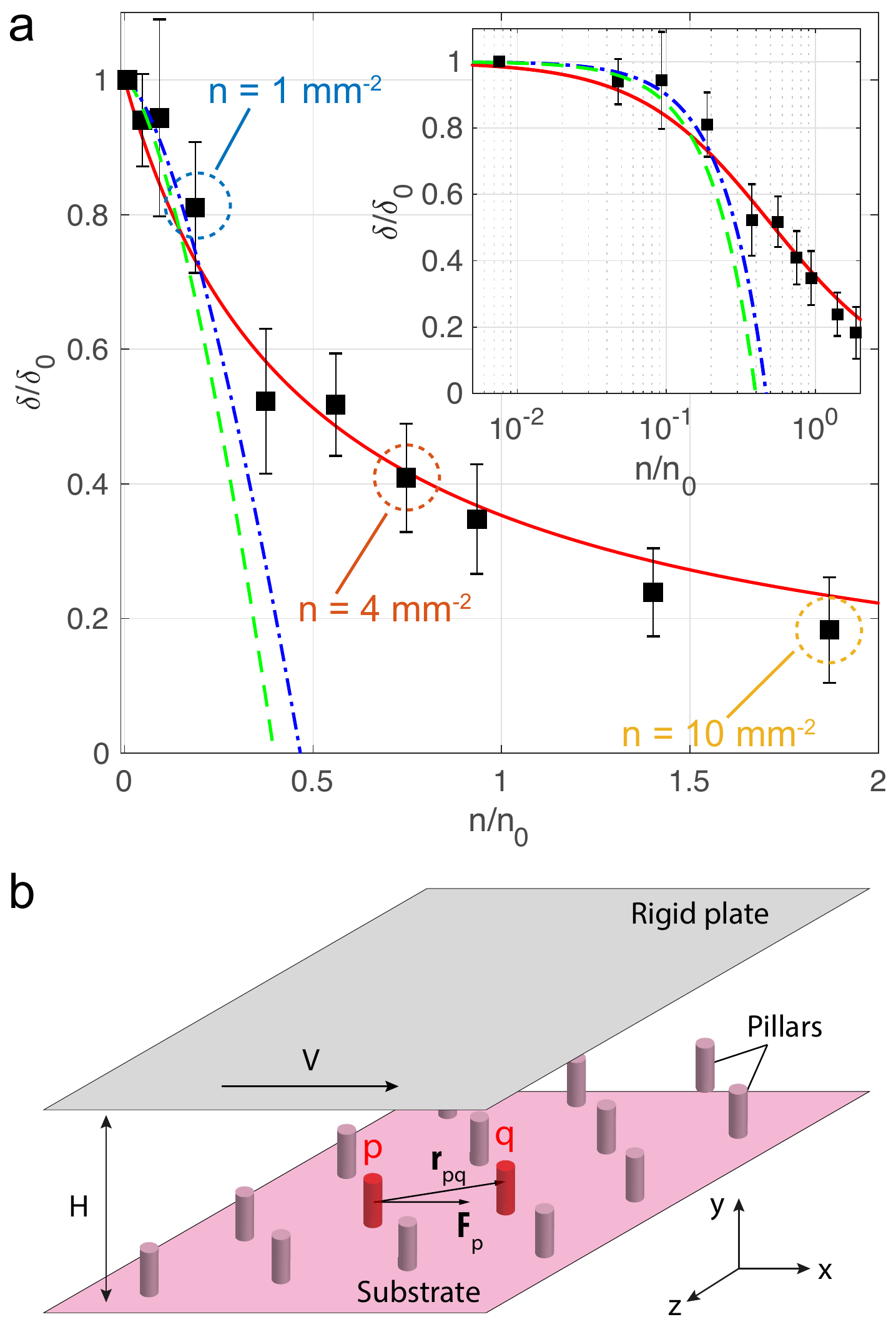}
\caption{(a) Dimensionless pillar tip deformation $\delta/\delta_0$ versus normalized pillars density $n/n_0$ with $n_0 = 5.3$~mm$^{-2}$ (squares). Each point is an average over 9 experiments and error bars are taken as the standard deviation over these 9 measurements. The green dashed (\textit{resp.} blue dash-dot) line is the prediction of Eq.~\ref{eq:deltadiscrete} with the the 5 nearest neighbors (\textit{resp.} Eq.~\ref{eqn:lauga-law-low-phi} in the continuum limit). The red solid curve is a fit with Eq.~\ref{eqn:result-alvarado} based on the phenomenological model of~\citep{AlvaradoHosoiNaturePhysics2017} where the value of $\alpha$ has been fitted. Points surrounded by dashed line circles correspond to those in Fig.~\ref{fig:fig2} with the same color code. Inset: Log-linear plot of the main graph. (b) Sketch of the hair assembly system with the geometrical characteristics used in the dilute model. The upper rigid plate positioned at a distance $H$ from the bottom of the pool is sheared along the $x$ direction with a velocity $\textbf{V}$.}
\label{fig:fig3}
\end{figure}

At all $n$, we still measure a linear relationship between $\delta$ and $\sigma$ with $\delta=\kappa \sigma$ (Fig.~\ref{fig:fig2}). Note that we normalized the deflection by $\kappa_0$ to take into account sample to sample variability. The slope of these normalized curves is then $\kappa/\kappa_0=\delta/\delta_0$. It is smaller than one and decreases with $n$, implying that denser pillars bend less than isolated ones.  Figure~\ref{fig:fig3}a shows, for all combined experiments, the dependence of $\delta/\delta_0$ with $n$ normalized by a characteristic density $n_0=1/L^2 \approx 5.3$~mm$^{-2}$. Note that the reference point at $n=0.04$~mm$^{-2}$ ($n/n_0=7. 10^{-3}$) has been added on the graph with $\delta/\delta_0=1$. Clearly, within error bars, $\delta/\delta_0$ decreases non-linearly with $n$.
    
Theoretically, the bending of an isolated elastic cylindrical pillar subject to a given flow has been derived in~\cite{Lauga2016}. In steady state, the balance between bending and the drag force from the fluid yields (in units of $L$)
\begin{equation}
\frac{d^4\delta}{dy^4} = -u_x(y)
\label{eq:laugamodel2}
\end{equation}
where $u_x$ is the flow velocity along the $x$ direction and $y$ the pillar's longitudinal coordinate (Fig.~\ref{fig:fig3}b). Equation~\ref{eq:laugamodel2} can be solved with boundary conditions $\delta(0)=0, \delta'(0)=0$ (clamped pillar at its base) and $\delta''(1)=0, \delta'''(1)=0$ (free pillars tips), where prime symbols stand for spatial derivatives with respect to $y$. For instance, for an isolated cylindrical pillar subjected to a shear flow, one obtains Eq.~\ref{eq:laugamodel1}. As a first attempt to model our data, we have derived an exact calculation of the induced flow perturbation by all pillars at the location of a given pillar. A pillar's deformation along the flow direction $\textbf{u}=\textbf{V}\,y/H$ (with $\textbf{V}$ the upper plate velocity, \textit{see} Fig.~\ref{fig:fig3}b) is due to the drag force from the fluid. In turn, the pillar exerts an opposite force with the same magnitude in the $-\textbf{u}$ direction. Consequently, this force tends to decrease the effective total flow in the $\textbf{u}$ direction acting on all other pillars. One thus expects a reduction of $\delta$ in pillar assemblies, \textit{i.e.} a \textit{collective stiffening}. Below, we first compute the flow field induced by a single pillar and then sum up all individual pillars contributions to obtain the net flow from the whole assembly.\\
\indent The flow induced by one pillar is calculated as resulting from the superposition of point forces along the pillar. For a point force located at a distance $h$ above the flat bottom surface, and assuming no-slip boundary conditions, the flow can be obtained using the hydrodynamic image method~\cite{Blake1971,BlakeChwang1974}. At a given location $\textbf{r}_{pq}$ from the point force $\textbf{F}_p$, the flow can be computed analytically in the dilute limit ($|\textbf{r}_{pq}|\gg L$) as

\begin{equation}
\textbf{u}_{p\rightarrow q}(\textbf{r}_{pq}) = \frac{3h}{4\pi\eta}\frac{\left(\textbf{F}_p\cdot \textbf{e}_{pq}\right)\textbf{e}_{pq}}{|\textbf{r}_{pq}|^3} y,
\label{Blake1:Eq}
\end{equation}
where $\textbf{e}_{pq}=\textbf{r}_{pq}/|\textbf{r}_{pq}|$. The whole net flow induced by one pillar $p$ can be computed by the superposition of the force per unit length on the fluid $\textbf{f}_p$ along the length of the pillar. In the limit of small deformations, it may be written as
\begin{equation}
\textbf{u}^{\text{net}}_{p\rightarrow q}(\textbf{r}_{pq}) = \frac{3}{4\pi\eta}\frac{\left(\int_0^L h \textbf{f}_p(h)\text{d}h\right)\cdot \textbf{e}_{pq}}{|\textbf{r}_{pq}|^3} y \textbf{e}_{pq}
\end{equation}    
Taking $\textbf{f}_p=-4\pi\eta u(h) \textbf{e}_x$ as a viscous force density and approximating the background flow as a shear flow $u(h)=\dot \gamma h$, one obtains
\begin{equation}
\textbf{u}^{\text{net}}_{p\rightarrow q}(\textbf{r}_{pq}) = -(\dot{\gamma} y )\frac{L^3}{|\textbf{r}_{pq}|^3}(\textbf{e}_x\cdot\textbf{e}_{pq})\ \textbf{e}_{pq}
\end{equation}

The collective deformation in the dilute limit is obtained by summing up on all pillars these hydrodynamic interactions. The total flow acting on pillar $q$ is therefore 
$\textbf{u}_q=\sum \limits_{p \ne q} \textbf{u}^{\text{net}}_{p\rightarrow q}(\textbf{r}_{pq})$. The deflection $\delta_q$ of the tip of pillar $q$ is obtained by solving Eq.~\ref{eq:laugamodel2} with the total flow given by the sum of the background flow plus perturbations, $\textbf{u}~+~\textbf{u}_q$, yielding 
\begin{equation}
\frac{\delta_q}{\delta_0} = \left|\textbf{e}_x-L^3 \sum_{p\ne q}\frac{1}{|\textbf{r}_{pq}|^3}(\textbf{e}_x\cdot\textbf{e}_{pq})\textbf{e}_{pq}\right|
\label{eq:deltadiscrete}
\end{equation}

The right hand side of Eq.~\ref{eq:deltadiscrete} can be computed analytically for both square and hexagonal lattices of pillars (\textit{see} Supplementary Material). In both cases,  we obtain that this term does not depend on the orientation of the lattice with respect to $\textbf{e}_x$. Alternatively, we may  estimate this term by taking the continuous limit. In cylindrical coordinates centered on pillar $p$, with $\textbf{e}_r$ and $\textbf{e}_{\theta}$ the unit vectors, $r$ the position of the pillar, Eq.~\ref{eq:deltadiscrete} becomes
\begin{equation}
\frac{\delta_q}{\delta_0} = \left|\textbf{e}_x-\int_0^{2\pi}\int_d^\infty n\frac{\cos{\theta}\textbf{e}_r}{r^3}rdrd\theta\right| = 1 - \pi \left(n L^2\right)^{3/2}
\label{eqn:lauga-law-low-phi}
\end{equation}

Predictions of Eq.~\ref{eq:deltadiscrete}, taking into account the five nearest neighbors to match with our experimental system, are plotted in Fig.~\ref{fig:fig3}a with the dashed line. The result from the continuous limit approximation of Eq.~\ref{eqn:lauga-law-low-phi} is plotted with the dot-dash line. The theoretical results agree with the experiments in the small $n/n_0$ limit, as expected from the dilute approximation. 

To go beyond our microscopic model valid in the dilute limit, we use the model of~\citet{AlvaradoHosoiNaturePhysics2017}. In that work, the authors computed the shear flow in a dense assembly of high aspect ratio cylindrical pillars treated as a porous medium of effective height $h_e$. Solving both Stokes' equation for the flow above the pillar bed and Brinkman's equation for the flow in the bed, they obtain the flow velocity $u(y)$ valid for $0<y<h_e$ as
\begin{equation}
u(y) = V \frac{\Delta \sinh(y/\Delta)}{(H-h_e)\cosh(h_e/\Delta)+\Delta \sinh(h_e/\Delta)}
\label{eqn:u_y_packed}
\end{equation}
where $V = \rho \omega$ is the velocity of the upper plate (Fig.~\ref{fig:fig3}b), $\Delta$ the pore size and $H$ the gap between the plates. For small deformations $\delta \ll L$, and thus $h_e\simeq L$. To compute pillars bending, we solved Eq.~\ref{eq:laugamodel2} with $u_x$ given by Eq.~\ref{eqn:u_y_packed}, yielding
\begin{equation}
\frac{\delta}{\delta_0} = \frac{120}{11}\frac{\displaystyle \beta^5S-\beta^4-\frac{1}{2}\beta^3S+\frac{1}{3}\beta^2C}{\displaystyle\left(1-L/H\right)C+\beta (L/H) S}
\label{eqn:result-alvarado}
\end{equation}
\noindent with $\beta = \Delta/L$, $S = \sinh(\beta^{-1})$ and $C = \cosh(\beta^{-1})$. Taking the pore size as $\Delta=\alpha/\sqrt{n}=\alpha d$, the data of Fig.~\ref{fig:fig3} have been fitted with Eq.~\ref{eqn:result-alvarado} with $\alpha$ as a free parameter. This model captures our data on the whole range of $n/n_0$ (Fig.~\ref{fig:fig3}a) with $\alpha=0.35\pm0.02$ and thus a pore size $\Delta$ of the order of the mesh size $d$. The parameter $\alpha$ relates the mesh size of the lattice to the pore size of the hair assembly. A similar value of 0.46 was obtained for turbulent flow over  carbon nanotube forests by Battiato \textit{et al.}~\cite{battiato2010}.

\indent Soft hair assemblies are ubiquitous in biology~\cite{du2019dynamics}. Recently, Pellicciotta \textit{et al.}~\cite{pellicciotta2020} have studied the collective beating of active motile cilia of brain cells subjected to oscillatory flows. They demonstrated  an enhanced hydrodynamic screening with the number of cilia which reduces their synchronization with the external flow. Similarly, in flagellar systems, pairs of beating flagella of unicellular micro-organisms can be synchronized solely through hydrodynamic interactions in the far field~\cite{brumley2014}. However, for  micro-organisms bearing few flagella, this synchronization is more complex and involves an elastic basal coupling in addition to hydrodynamics interactions~\cite{wan2016}. Overall, there is however a lack of quantitative studies of how dense assemblies of pillars deform in  flows. \citet{coq2011collective} investigated the dynamics of a bed of magnetic micro-cilia distributed on a square lattice and submitted to a precessing magnetic field. They showed that at high surface density, the collective beating yields a symmetry breaking of the circular precession. Like us, they interpreted their results using the models of~Refs.~\cite{Blake1971,BlakeChwang1974}. They derived an expression for the dependence of the amplitude of the hydrodynamic interactions versus density, which is very similar to our Eq.~\ref{eq:deltadiscrete} (\textit{see} Supplemental Material of~\cite{coq2011collective}). However, they do not test experimentally its dependence with the pillars density. Moreover, it is only valid at low pillars density when $n \ll 1/L^2$ and thus cannot apply to their experimental results for which $n L^2 \approx 20$. In another context, Bhushan~\cite{bhushan2011} has also reported drag reductions in microtextured channels with undeformable pillars. However, the reduction results from super-hydrophobicity effects and not hydrodynamic interactions.
\\
\indent Our study provides a complete experimental test of this model over a wide range of $n$. We demonstrate that the bending of pillars remains proportional to the shear stress at any $n$. We also show that in the dilute limit, our microscopic model is in reasonable agreement with our data. The collective bending of the hair assemblies decays typically over a density of order $1/L^2$. In addition, we demonstrate that hydrodynamical interactions do not depend on the topology of the lattices (square or hexagonal) nor on its orientation with respect to the flow. At higher $n$, typically $n L^2 \gtrsim 0.5$, our data show discrepancies with the dilute model. This could result from the fact that, to compute Eq.~\ref{Blake1:Eq}, we neglected the near-field terms in $1/r^5$ which are expected to have important contributions at large $n$. To describe our measurements over the whole pillar density range, we have thus used the recent phenomenological model of~\cite{AlvaradoHosoiNaturePhysics2017} describing the pillar assemblies as a porous medium. In~\cite{AlvaradoHosoiNaturePhysics2017}, the authors have successfully tested their model, however limited to three different densities (all above $n_0$), and they did not probe the deflections of pillars induced by such flows. We show here that solving Eq.~\ref{eq:laugamodel2} with their flow field allows to reproduce faithfully the pillars deflections at any density with the porosity as a single fit parameter. We find again that the characteristic density is of order $1/L^2$. This analytical method complements the very recent numerical predictions of Stein and Shelley~\cite{stein2019coarse}. Interestingly, an asymptotic expansion of Eq.~\ref{eqn:result-alvarado} at very high density ($n \gg n_0$) yields $\delta/\delta_0 \sim 1/n$, in agreement with~\cite{stein2019coarse}.\\
\indent In many biological systems, the deformation of soft hair assemblies is the primary mechanical input measured by mechanoreceptors. We have shown that at high density of pillars, their effective deformation can be significantly reduced. In the particular case of mammalian tongues for instance, filiform papillae are densely packed with a typical density $n L^2$ that ranges from 0.1 to 1. Our work shows that at these densities, the reduction of papillae deflections should reach about half of their nominal deflection if papillae were isolated. Biologically, this mechanism could induce an enhanced protection of the sensory structure by avoiding large deformations. We have also shown that this collective stiffening is independent of spatial organization and orientation of the lattice with respect to the flow. It therefore suggests that such sensory systems are robust to flow direction. Beyond the particular case of vertebrate tongues, our results should be also applicable to a wide range of biological systems.\\

\begin{acknowledgments}
The authors acknowledge financial support from Centre de Recherche et D\'eveloppement Nestl\'e S.A.S., Marne la Vall\'ee, France and Nestec Ltd, Vevey, Switzerland. They also thank I.~Barbotteau and G.~Marchesini (Nestl\'{e} Dairy Center, Lisieux, France) for their careful reading of the manuscript and support. Finally, the authors acknowledge the support of C.~J. Pipe (Nestl\'{e} Research Center, Switzerland) and thank G.~Debr\'egeas (Laboratoire Jean Perrin, Paris, France) and C.~Fr\'etigny (Laboratoire SIMM, Paris, France) for fruitful discussions.
\end{acknowledgments}


\begin{thebibliography}{31}%
\makeatletter
\providecommand \@ifxundefined [1]{%
 \@ifx{#1\undefined}
}%
\providecommand \@ifnum [1]{%
 \ifnum #1\expandafter \@firstoftwo
 \else \expandafter \@secondoftwo
 \fi
}%
\providecommand \@ifx [1]{%
 \ifx #1\expandafter \@firstoftwo
 \else \expandafter \@secondoftwo
 \fi
}%
\providecommand \natexlab [1]{#1}%
\providecommand \enquote  [1]{``#1''}%
\providecommand \bibnamefont  [1]{#1}%
\providecommand \bibfnamefont [1]{#1}%
\providecommand \citenamefont [1]{#1}%
\providecommand \href@noop [0]{\@secondoftwo}%
\providecommand \href [0]{\begingroup \@sanitize@url \@href}%
\providecommand \@href[1]{\@@startlink{#1}\@@href}%
\providecommand \@@href[1]{\endgroup#1\@@endlink}%
\providecommand \@sanitize@url [0]{\catcode `\\12\catcode `\$12\catcode
  `\&12\catcode `\#12\catcode `\^12\catcode `\_12\catcode `\%12\relax}%
\providecommand \@@startlink[1]{}%
\providecommand \@@endlink[0]{}%
\providecommand \url  [0]{\begingroup\@sanitize@url \@url }%
\providecommand \@url [1]{\endgroup\@href {#1}{\urlprefix }}%
\providecommand \urlprefix  [0]{URL }%
\providecommand \Eprint [0]{\href }%
\providecommand \doibase [0]{http://dx.doi.org/}%
\providecommand \selectlanguage [0]{\@gobble}%
\providecommand \bibinfo  [0]{\@secondoftwo}%
\providecommand \bibfield  [0]{\@secondoftwo}%
\providecommand \translation [1]{[#1]}%
\providecommand \BibitemOpen [0]{}%
\providecommand \bibitemStop [0]{}%
\providecommand \bibitemNoStop [0]{.\EOS\space}%
\providecommand \EOS [0]{\spacefactor3000\relax}%
\providecommand \BibitemShut  [1]{\csname bibitem#1\endcsname}%
\let\auto@bib@innerbib\@empty
%</preamble>
\bibitem [{\citenamefont {Delmas}\ \emph {et~al.}(2011)\citenamefont {Delmas},
  \citenamefont {Hao},\ and\ \citenamefont
  {Rodat-Despoix}}]{delmas2011molecular}%
  \BibitemOpen
  \bibfield  {author} {\bibinfo {author} {\bibfnamefont {P.}~\bibnamefont
  {Delmas}}, \bibinfo {author} {\bibfnamefont {J.}~\bibnamefont {Hao}}, \ and\
  \bibinfo {author} {\bibfnamefont {L.}~\bibnamefont {Rodat-Despoix}},\
  }\href@noop {} {\bibfield  {journal} {\bibinfo  {journal} {Nature Reviews
  Neuroscience}\ }\textbf {\bibinfo {volume} {12}},\ \bibinfo {pages} {139}
  (\bibinfo {year} {2011})}\BibitemShut {NoStop}%
\bibitem [{\citenamefont {Arabzadeh}\ \emph {et~al.}(2016)\citenamefont
  {Arabzadeh}, \citenamefont {von Heimendahl},\ and\ \citenamefont
  {Diamond}}]{arabzadeh2016vibrissal}%
  \BibitemOpen
  \bibfield  {author} {\bibinfo {author} {\bibfnamefont {E.}~\bibnamefont
  {Arabzadeh}}, \bibinfo {author} {\bibfnamefont {M.}~\bibnamefont {von
  Heimendahl}}, \ and\ \bibinfo {author} {\bibfnamefont {M.}~\bibnamefont
  {Diamond}},\ }in\ \href@noop {} {\emph {\bibinfo {booktitle} {Scholarpedia of
  Touch}}}\ (\bibinfo  {publisher} {Springer},\ \bibinfo {year} {2016})\ pp.\
  \bibinfo {pages} {737--749}\BibitemShut {NoStop}%
\bibitem [{\citenamefont {Bathellier}\ \emph {et~al.}(2005)\citenamefont
  {Bathellier}, \citenamefont {Barth}, \citenamefont {Albert},\ and\
  \citenamefont {Humphrey}}]{bathellier2005viscosity}%
  \BibitemOpen
  \bibfield  {author} {\bibinfo {author} {\bibfnamefont {B.}~\bibnamefont
  {Bathellier}}, \bibinfo {author} {\bibfnamefont {F.~G.}\ \bibnamefont
  {Barth}}, \bibinfo {author} {\bibfnamefont {J.~T.}\ \bibnamefont {Albert}}, \
  and\ \bibinfo {author} {\bibfnamefont {J.~A.}\ \bibnamefont {Humphrey}},\
  }\href@noop {} {\bibfield  {journal} {\bibinfo  {journal} {Journal of
  comparative physiology A}\ }\textbf {\bibinfo {volume} {191}},\ \bibinfo
  {pages} {733} (\bibinfo {year} {2005})}\BibitemShut {NoStop}%
\bibitem [{\citenamefont {Dangles}\ \emph {et~al.}(2006)\citenamefont
  {Dangles}, \citenamefont {Pierre}, \citenamefont {Magal}, \citenamefont
  {Vannier},\ and\ \citenamefont {Casas}}]{dangles2006ontogeny}%
  \BibitemOpen
  \bibfield  {author} {\bibinfo {author} {\bibfnamefont {O.}~\bibnamefont
  {Dangles}}, \bibinfo {author} {\bibfnamefont {D.}~\bibnamefont {Pierre}},
  \bibinfo {author} {\bibfnamefont {C.}~\bibnamefont {Magal}}, \bibinfo
  {author} {\bibfnamefont {F.}~\bibnamefont {Vannier}}, \ and\ \bibinfo
  {author} {\bibfnamefont {J.}~\bibnamefont {Casas}},\ }\href@noop {}
  {\bibfield  {journal} {\bibinfo  {journal} {Journal of Experimental Biology}\
  }\textbf {\bibinfo {volume} {209}},\ \bibinfo {pages} {4363} (\bibinfo {year}
  {2006})}\BibitemShut {NoStop}%
\bibitem [{\citenamefont {Chagnaud}\ \emph {et~al.}(2008)\citenamefont
  {Chagnaud}, \citenamefont {Br{\"u}cker}, \citenamefont {Hofmann},\ and\
  \citenamefont {Bleckmann}}]{chagnaud2008measuring}%
  \BibitemOpen
  \bibfield  {author} {\bibinfo {author} {\bibfnamefont {B.~P.}\ \bibnamefont
  {Chagnaud}}, \bibinfo {author} {\bibfnamefont {C.}~\bibnamefont
  {Br{\"u}cker}}, \bibinfo {author} {\bibfnamefont {M.~H.}\ \bibnamefont
  {Hofmann}}, \ and\ \bibinfo {author} {\bibfnamefont {H.}~\bibnamefont
  {Bleckmann}},\ }\href@noop {} {\bibfield  {journal} {\bibinfo  {journal}
  {Journal of Neuroscience}\ }\textbf {\bibinfo {volume} {28}},\ \bibinfo
  {pages} {4479} (\bibinfo {year} {2008})}\BibitemShut {NoStop}%
\bibitem [{\citenamefont {Ghom}\ and\ \citenamefont
  {Mhaske}(2008)}]{ghom2008textbook}%
  \BibitemOpen
  \bibfield  {author} {\bibinfo {author} {\bibfnamefont {A.}~\bibnamefont
  {Ghom}}\ and\ \bibinfo {author} {\bibfnamefont {S.}~\bibnamefont {Mhaske}},\
  }\href@noop {} {\emph {\bibinfo {title} {Textbook of oral pathology}}}\
  (\bibinfo  {publisher} {Jaypee Brothers Medical Publishers New Delhi},\
  \bibinfo {year} {2008})\BibitemShut {NoStop}%
\bibitem [{\citenamefont {Yamashita}\ and\ \citenamefont
  {OdDalkhsuren}(2012)}]{BookShimizu2012}%
  \BibitemOpen
  \bibfield  {author} {\bibinfo {author} {\bibfnamefont {K.}~\bibnamefont
  {Yamashita}}\ and\ \bibinfo {author} {\bibfnamefont {S.}~\bibnamefont
  {OdDalkhsuren}},\ }in\ \href@noop {} {\emph {\bibinfo {booktitle} {Tongue:
  Anatomy, Kinematics and Diseases}}},\ \bibinfo {editor} {edited by\ \bibinfo
  {editor} {\bibfnamefont {H.}~\bibnamefont {Kat}}\ and\ \bibinfo {editor}
  {\bibfnamefont {T.}~\bibnamefont {Shimizu}}}\ (\bibinfo  {publisher} {Nova
  Science Publishers, Inc.},\ \bibinfo {address} {New-York},\ \bibinfo {year}
  {2012})\ pp.\ \bibinfo {pages} {143--154}\BibitemShut {NoStop}%
\bibitem [{\citenamefont {Doty}(2015)}]{doty2015handbook}%
  \BibitemOpen
  \bibfield  {author} {\bibinfo {author} {\bibfnamefont {R.~L.}\ \bibnamefont
  {Doty}},\ }\href@noop {} {\emph {\bibinfo {title} {Handbook of olfaction and
  gustation}}}\ (\bibinfo  {publisher} {John Wiley \& Sons},\ \bibinfo {year}
  {2015})\BibitemShut {NoStop}%
\bibitem [{\citenamefont {Malicki}\ and\ \citenamefont
  {Johnson}(2017)}]{malicki2017cilium}%
  \BibitemOpen
  \bibfield  {author} {\bibinfo {author} {\bibfnamefont {J.~J.}\ \bibnamefont
  {Malicki}}\ and\ \bibinfo {author} {\bibfnamefont {C.~A.}\ \bibnamefont
  {Johnson}},\ }\href@noop {} {\bibfield  {journal} {\bibinfo  {journal}
  {Trends in Cell Biology}\ }\textbf {\bibinfo {volume} {27}},\ \bibinfo
  {pages} {126} (\bibinfo {year} {2017})}\BibitemShut {NoStop}%
\bibitem [{\citenamefont {Quist}\ \emph {et~al.}(2014)\citenamefont {Quist},
  \citenamefont {Seghete}, \citenamefont {Huet}, \citenamefont {Murphey},\ and\
  \citenamefont {Hartmann}}]{quist2014modeling}%
  \BibitemOpen
  \bibfield  {author} {\bibinfo {author} {\bibfnamefont {B.~W.}\ \bibnamefont
  {Quist}}, \bibinfo {author} {\bibfnamefont {V.}~\bibnamefont {Seghete}},
  \bibinfo {author} {\bibfnamefont {L.~A.}\ \bibnamefont {Huet}}, \bibinfo
  {author} {\bibfnamefont {T.~D.}\ \bibnamefont {Murphey}}, \ and\ \bibinfo
  {author} {\bibfnamefont {M.~J.}\ \bibnamefont {Hartmann}},\ }\href@noop {}
  {\bibfield  {journal} {\bibinfo  {journal} {Journal of Neuroscience}\
  }\textbf {\bibinfo {volume} {34}},\ \bibinfo {pages} {9828} (\bibinfo {year}
  {2014})}\BibitemShut {NoStop}%
\bibitem [{\citenamefont {Boubenec}\ \emph {et~al.}(2012)\citenamefont
  {Boubenec}, \citenamefont {Shulz},\ and\ \citenamefont
  {Debr{\'e}geas}}]{boubenec2012whisker}%
  \BibitemOpen
  \bibfield  {author} {\bibinfo {author} {\bibfnamefont {Y.}~\bibnamefont
  {Boubenec}}, \bibinfo {author} {\bibfnamefont {D.~E.}\ \bibnamefont {Shulz}},
  \ and\ \bibinfo {author} {\bibfnamefont {G.}~\bibnamefont {Debr{\'e}geas}},\
  }\href@noop {} {\bibfield  {journal} {\bibinfo  {journal} {Frontiers in
  behavioral neuroscience}\ }\textbf {\bibinfo {volume} {6}},\ \bibinfo {pages}
  {74} (\bibinfo {year} {2012})}\BibitemShut {NoStop}%
\bibitem [{\citenamefont {Venier}\ \emph {et~al.}(1994)\citenamefont {Venier},
  \citenamefont {Maggs}, \citenamefont {Carlier},\ and\ \citenamefont
  {Pantaloni}}]{Venier1994}%
  \BibitemOpen
  \bibfield  {author} {\bibinfo {author} {\bibfnamefont {P.}~\bibnamefont
  {Venier}}, \bibinfo {author} {\bibfnamefont {A.~C.}\ \bibnamefont {Maggs}},
  \bibinfo {author} {\bibfnamefont {M.~F.}\ \bibnamefont {Carlier}}, \ and\
  \bibinfo {author} {\bibfnamefont {D.}~\bibnamefont {Pantaloni}},\ }\href@noop
  {} {\bibfield  {journal} {\bibinfo  {journal} {Journal of Biological
  Chemistry}\ }\textbf {\bibinfo {volume} {269}},\ \bibinfo {pages} {13353}
  (\bibinfo {year} {1994})}\BibitemShut {NoStop}%
\bibitem [{\citenamefont {Lauga}\ \emph {et~al.}(2016)\citenamefont {Lauga},
  \citenamefont {Pipe},\ and\ \citenamefont {{Le
  R{\'{e}}v{\'{e}}rend}}}]{Lauga2016}%
  \BibitemOpen
  \bibfield  {author} {\bibinfo {author} {\bibfnamefont {E.}~\bibnamefont
  {Lauga}}, \bibinfo {author} {\bibfnamefont {C.~J.}\ \bibnamefont {Pipe}}, \
  and\ \bibinfo {author} {\bibfnamefont {B.}~\bibnamefont {{Le
  R{\'{e}}v{\'{e}}rend}}},\ }\href@noop {} {\bibfield  {journal} {\bibinfo
  {journal} {Frontiers in Physics}\ }\textbf {\bibinfo {volume} {4}},\ \bibinfo
  {pages} {35} (\bibinfo {year} {2016})}\BibitemShut {NoStop}%
\bibitem [{\citenamefont {du~Roure}\ \emph {et~al.}(2019)\citenamefont
  {du~Roure}, \citenamefont {Lindner}, \citenamefont {Nazockdast},\ and\
  \citenamefont {Shelley}}]{du2019dynamics}%
  \BibitemOpen
  \bibfield  {author} {\bibinfo {author} {\bibfnamefont {O.}~\bibnamefont
  {du~Roure}}, \bibinfo {author} {\bibfnamefont {A.}~\bibnamefont {Lindner}},
  \bibinfo {author} {\bibfnamefont {E.~N.}\ \bibnamefont {Nazockdast}}, \ and\
  \bibinfo {author} {\bibfnamefont {M.~J.}\ \bibnamefont {Shelley}},\
  }\href@noop {} {\bibfield  {journal} {\bibinfo  {journal} {Annual Review of
  Fluid Mechanics}\ }\textbf {\bibinfo {volume} {51}},\ \bibinfo {pages} {539}
  (\bibinfo {year} {2019})}\BibitemShut {NoStop}%
\bibitem [{\citenamefont {Claverie}\ \emph {et~al.}(2017)\citenamefont
  {Claverie}, \citenamefont {Boubenec}, \citenamefont {Debr{\'e}geas},
  \citenamefont {Prevost},\ and\ \citenamefont
  {Wandersman}}]{claverie2017whisker}%
  \BibitemOpen
  \bibfield  {author} {\bibinfo {author} {\bibfnamefont {L.~N.}\ \bibnamefont
  {Claverie}}, \bibinfo {author} {\bibfnamefont {Y.}~\bibnamefont {Boubenec}},
  \bibinfo {author} {\bibfnamefont {G.}~\bibnamefont {Debr{\'e}geas}}, \bibinfo
  {author} {\bibfnamefont {A.~M.}\ \bibnamefont {Prevost}}, \ and\ \bibinfo
  {author} {\bibfnamefont {E.}~\bibnamefont {Wandersman}},\ }\href@noop {}
  {\bibfield  {journal} {\bibinfo  {journal} {Frontiers in behavioral
  neuroscience}\ }\textbf {\bibinfo {volume} {10}},\ \bibinfo {pages} {251}
  (\bibinfo {year} {2017})}\BibitemShut {NoStop}%
\bibitem [{\citenamefont {Axtmann}\ \emph {et~al.}(2016)\citenamefont
  {Axtmann}, \citenamefont {Hegner}, \citenamefont {Br{\"u}cker},\ and\
  \citenamefont {Rist}}]{axtmann2016investigation}%
  \BibitemOpen
  \bibfield  {author} {\bibinfo {author} {\bibfnamefont {G.}~\bibnamefont
  {Axtmann}}, \bibinfo {author} {\bibfnamefont {F.}~\bibnamefont {Hegner}},
  \bibinfo {author} {\bibfnamefont {C.}~\bibnamefont {Br{\"u}cker}}, \ and\
  \bibinfo {author} {\bibfnamefont {U.}~\bibnamefont {Rist}},\ }\href@noop {}
  {\bibfield  {journal} {\bibinfo  {journal} {Journal of Fluids and
  Structures}\ }\textbf {\bibinfo {volume} {66}},\ \bibinfo {pages} {110}
  (\bibinfo {year} {2016})}\BibitemShut {NoStop}%
\bibitem [{\citenamefont {Wexler}\ \emph {et~al.}(2013)\citenamefont {Wexler},
  \citenamefont {Trinh}, \citenamefont {Berthet}, \citenamefont {Quennouz},
  \citenamefont {du~Roure}, \citenamefont {Huppert}, \citenamefont {Lindner},\
  and\ \citenamefont {Stone}}]{wexler2013bending}%
  \BibitemOpen
  \bibfield  {author} {\bibinfo {author} {\bibfnamefont {J.~S.}\ \bibnamefont
  {Wexler}}, \bibinfo {author} {\bibfnamefont {P.~H.}\ \bibnamefont {Trinh}},
  \bibinfo {author} {\bibfnamefont {H.}~\bibnamefont {Berthet}}, \bibinfo
  {author} {\bibfnamefont {N.}~\bibnamefont {Quennouz}}, \bibinfo {author}
  {\bibfnamefont {O.}~\bibnamefont {du~Roure}}, \bibinfo {author}
  {\bibfnamefont {H.~E.}\ \bibnamefont {Huppert}}, \bibinfo {author}
  {\bibfnamefont {A.}~\bibnamefont {Lindner}}, \ and\ \bibinfo {author}
  {\bibfnamefont {H.~A.}\ \bibnamefont {Stone}},\ }\href@noop {} {\bibfield
  {journal} {\bibinfo  {journal} {Journal of fluid mechanics}\ }\textbf
  {\bibinfo {volume} {720}},\ \bibinfo {pages} {517} (\bibinfo {year}
  {2013})}\BibitemShut {NoStop}%
\bibitem [{\citenamefont {Thomazo}\ \emph {et~al.}(2019)\citenamefont
  {Thomazo}, \citenamefont {Contreras~Pastenes}, \citenamefont {Pipe},
  \citenamefont {Le~R\'ev\'erend}, \citenamefont {Wandersman},\ and\
  \citenamefont {Prevost}}]{ThomazoJRSI2019}%
  \BibitemOpen
  \bibfield  {author} {\bibinfo {author} {\bibfnamefont {J.-B.}\ \bibnamefont
  {Thomazo}}, \bibinfo {author} {\bibfnamefont {J.}~\bibnamefont
  {Contreras~Pastenes}}, \bibinfo {author} {\bibfnamefont {C.~J.}\ \bibnamefont
  {Pipe}}, \bibinfo {author} {\bibfnamefont {B.}~\bibnamefont
  {Le~R\'ev\'erend}}, \bibinfo {author} {\bibfnamefont {E.}~\bibnamefont
  {Wandersman}}, \ and\ \bibinfo {author} {\bibfnamefont {A.~M.}\ \bibnamefont
  {Prevost}},\ }\href@noop {} {\bibfield  {journal} {\bibinfo  {journal}
  {Journal of the Royal Society Interface}\ }\textbf {\bibinfo {volume} {16}},\
  \bibinfo {pages} {20190362} (\bibinfo {year} {2019})}\BibitemShut {NoStop}%
\bibitem [{\citenamefont {Lauga}\ and\ \citenamefont
  {Powers}(2009)}]{lauga2009hydrodynamics}%
  \BibitemOpen
  \bibfield  {author} {\bibinfo {author} {\bibfnamefont {E.}~\bibnamefont
  {Lauga}}\ and\ \bibinfo {author} {\bibfnamefont {T.~R.}\ \bibnamefont
  {Powers}},\ }\href@noop {} {\bibfield  {journal} {\bibinfo  {journal}
  {Reports on Progress in Physics}\ }\textbf {\bibinfo {volume} {72}},\
  \bibinfo {pages} {096601} (\bibinfo {year} {2009})}\BibitemShut {NoStop}%
\bibitem [{\citenamefont {Weihs}(1973)}]{weihs1973hydromechanics}%
  \BibitemOpen
  \bibfield  {author} {\bibinfo {author} {\bibfnamefont {D.}~\bibnamefont
  {Weihs}},\ }\href@noop {} {\bibfield  {journal} {\bibinfo  {journal}
  {Nature}\ }\textbf {\bibinfo {volume} {241}},\ \bibinfo {pages} {290}
  (\bibinfo {year} {1973})}\BibitemShut {NoStop}%
\bibitem [{\citenamefont {Higdon}\ and\ \citenamefont
  {Corrsin}(1978)}]{higdon1978induced}%
  \BibitemOpen
  \bibfield  {author} {\bibinfo {author} {\bibfnamefont {J.}~\bibnamefont
  {Higdon}}\ and\ \bibinfo {author} {\bibfnamefont {S.}~\bibnamefont
  {Corrsin}},\ }\href@noop {} {\bibfield  {journal} {\bibinfo  {journal} {The
  American Naturalist}\ }\textbf {\bibinfo {volume} {112}},\ \bibinfo {pages}
  {727} (\bibinfo {year} {1978})}\BibitemShut {NoStop}%
\bibitem [{\citenamefont {Alvarado}\ \emph {et~al.}(2017)\citenamefont
  {Alvarado}, \citenamefont {Comtet}, \citenamefont {de~Langre},\ and\
  \citenamefont {Hosoi}}]{AlvaradoHosoiNaturePhysics2017}%
  \BibitemOpen
  \bibfield  {author} {\bibinfo {author} {\bibfnamefont {J.}~\bibnamefont
  {Alvarado}}, \bibinfo {author} {\bibfnamefont {J.}~\bibnamefont {Comtet}},
  \bibinfo {author} {\bibfnamefont {E.}~\bibnamefont {de~Langre}}, \ and\
  \bibinfo {author} {\bibfnamefont {A.~E.}\ \bibnamefont {Hosoi}},\ }\href@noop
  {} {\bibfield  {journal} {\bibinfo  {journal} {Nature Physics}\ }\textbf
  {\bibinfo {volume} {13}},\ \bibinfo {pages} {1014} (\bibinfo {year}
  {2017})}\BibitemShut {NoStop}%
\bibitem [{\citenamefont {Blake}(1971)}]{Blake1971}%
  \BibitemOpen
  \bibfield  {author} {\bibinfo {author} {\bibfnamefont {J.~R.}\ \bibnamefont
  {Blake}},\ }\href@noop {} {\bibfield  {journal} {\bibinfo  {journal}
  {Mathematical Proceedings of the Cambridge Philosophical Society}\ }\textbf
  {\bibinfo {volume} {70}},\ \bibinfo {pages} {303} (\bibinfo {year}
  {1971})}\BibitemShut {NoStop}%
\bibitem [{\citenamefont {Blake}\ and\ \citenamefont
  {Chwang}(1974)}]{BlakeChwang1974}%
  \BibitemOpen
  \bibfield  {author} {\bibinfo {author} {\bibfnamefont {J.~R.}\ \bibnamefont
  {Blake}}\ and\ \bibinfo {author} {\bibfnamefont {A.~T.}\ \bibnamefont
  {Chwang}},\ }\href@noop {} {\bibfield  {journal} {\bibinfo  {journal}
  {Journal of Engineering Mathematics}\ }\textbf {\bibinfo {volume} {8}},\
  \bibinfo {pages} {23} (\bibinfo {year} {1974})}\BibitemShut {NoStop}%
\bibitem [{\citenamefont {Battiato}\ \emph {et~al.}(2010)\citenamefont
  {Battiato}, \citenamefont {Bandaru},\ and\ \citenamefont
  {Tartakovsky}}]{battiato2010}%
  \BibitemOpen
  \bibfield  {author} {\bibinfo {author} {\bibfnamefont {I.}~\bibnamefont
  {Battiato}}, \bibinfo {author} {\bibfnamefont {P.~R.}\ \bibnamefont
  {Bandaru}}, \ and\ \bibinfo {author} {\bibfnamefont {D.~M.}\ \bibnamefont
  {Tartakovsky}},\ }\href@noop {} {\bibfield  {journal} {\bibinfo  {journal}
  {Physical review letters}\ }\textbf {\bibinfo {volume} {105}},\ \bibinfo
  {pages} {144504} (\bibinfo {year} {2010})}\BibitemShut {NoStop}%
\bibitem [{\citenamefont {Pellicciotta}\ \emph {et~al.}(2020)\citenamefont
  {Pellicciotta}, \citenamefont {Hamilton}, \citenamefont {Kotar},
  \citenamefont {Faucourt}, \citenamefont {Delgehyr}, \citenamefont {Spassky},\
  and\ \citenamefont {Cicuta}}]{pellicciotta2020}%
  \BibitemOpen
  \bibfield  {author} {\bibinfo {author} {\bibfnamefont {N.}~\bibnamefont
  {Pellicciotta}}, \bibinfo {author} {\bibfnamefont {E.}~\bibnamefont
  {Hamilton}}, \bibinfo {author} {\bibfnamefont {J.}~\bibnamefont {Kotar}},
  \bibinfo {author} {\bibfnamefont {M.}~\bibnamefont {Faucourt}}, \bibinfo
  {author} {\bibfnamefont {N.}~\bibnamefont {Delgehyr}}, \bibinfo {author}
  {\bibfnamefont {N.}~\bibnamefont {Spassky}}, \ and\ \bibinfo {author}
  {\bibfnamefont {P.}~\bibnamefont {Cicuta}},\ }\href@noop {} {\bibfield
  {journal} {\bibinfo  {journal} {Proceedings of the National Academy of
  Sciences}\ }\textbf {\bibinfo {volume} {117}},\ \bibinfo {pages} {8315}
  (\bibinfo {year} {2020})}\BibitemShut {NoStop}%
\bibitem [{\citenamefont {Brumley}\ \emph {et~al.}(2014)\citenamefont
  {Brumley}, \citenamefont {Wan}, \citenamefont {Polin},\ and\ \citenamefont
  {Goldstein}}]{brumley2014}%
  \BibitemOpen
  \bibfield  {author} {\bibinfo {author} {\bibfnamefont {D.~R.}\ \bibnamefont
  {Brumley}}, \bibinfo {author} {\bibfnamefont {K.~Y.}\ \bibnamefont {Wan}},
  \bibinfo {author} {\bibfnamefont {M.}~\bibnamefont {Polin}}, \ and\ \bibinfo
  {author} {\bibfnamefont {R.~E.}\ \bibnamefont {Goldstein}},\ }\href@noop {}
  {\bibfield  {journal} {\bibinfo  {journal} {Elife}\ }\textbf {\bibinfo
  {volume} {3}},\ \bibinfo {pages} {e02750} (\bibinfo {year}
  {2014})}\BibitemShut {NoStop}%
\bibitem [{\citenamefont {Wan}\ and\ \citenamefont
  {Goldstein}(2016)}]{wan2016}%
  \BibitemOpen
  \bibfield  {author} {\bibinfo {author} {\bibfnamefont {K.~Y.}\ \bibnamefont
  {Wan}}\ and\ \bibinfo {author} {\bibfnamefont {R.~E.}\ \bibnamefont
  {Goldstein}},\ }\href@noop {} {\bibfield  {journal} {\bibinfo  {journal}
  {Proceedings of the National Academy of Sciences}\ }\textbf {\bibinfo
  {volume} {113}},\ \bibinfo {pages} {E2784} (\bibinfo {year}
  {2016})}\BibitemShut {NoStop}%
\bibitem [{\citenamefont {Coq}\ \emph {et~al.}(2011)\citenamefont {Coq},
  \citenamefont {Bricard}, \citenamefont {Delapierre}, \citenamefont
  {Malaquin}, \citenamefont {du~Roure}, \citenamefont {Fermigier},\ and\
  \citenamefont {Bartolo}}]{coq2011collective}%
  \BibitemOpen
  \bibfield  {author} {\bibinfo {author} {\bibfnamefont {N.}~\bibnamefont
  {Coq}}, \bibinfo {author} {\bibfnamefont {A.}~\bibnamefont {Bricard}},
  \bibinfo {author} {\bibfnamefont {F.-D.}\ \bibnamefont {Delapierre}},
  \bibinfo {author} {\bibfnamefont {L.}~\bibnamefont {Malaquin}}, \bibinfo
  {author} {\bibfnamefont {O.}~\bibnamefont {du~Roure}}, \bibinfo {author}
  {\bibfnamefont {M.}~\bibnamefont {Fermigier}}, \ and\ \bibinfo {author}
  {\bibfnamefont {D.}~\bibnamefont {Bartolo}},\ }\href@noop {} {\bibfield
  {journal} {\bibinfo  {journal} {Physical Review Letters}\ }\textbf {\bibinfo
  {volume} {107}},\ \bibinfo {pages} {014501} (\bibinfo {year}
  {2011})}\BibitemShut {NoStop}%
\bibitem [{\citenamefont {Bhushan}(2011)}]{bhushan2011}%
  \BibitemOpen
  \bibfield  {author} {\bibinfo {author} {\bibfnamefont {B.}~\bibnamefont
  {Bhushan}},\ }\href@noop {} {\bibfield  {journal} {\bibinfo  {journal}
  {Beilstein journal of nanotechnology}\ }\textbf {\bibinfo {volume} {2}},\
  \bibinfo {pages} {66} (\bibinfo {year} {2011})}\BibitemShut {NoStop}%
\bibitem [{\citenamefont {Stein}\ and\ \citenamefont
  {Shelley}(2019)}]{stein2019coarse}%
  \BibitemOpen
  \bibfield  {author} {\bibinfo {author} {\bibfnamefont {D.~B.}\ \bibnamefont
  {Stein}}\ and\ \bibinfo {author} {\bibfnamefont {M.~J.}\ \bibnamefont
  {Shelley}},\ }\href@noop {} {\bibfield  {journal} {\bibinfo  {journal}
  {Physical Review Fluids}\ }\textbf {\bibinfo {volume} {4}},\ \bibinfo {pages}
  {073302} (\bibinfo {year} {2019})}\BibitemShut {NoStop}%
\end{thebibliography}
\end{document}